\documentclass[reprint,english,superscriptaddress,preprintnumbers,amsmath,amssymb,aps,prd]{revtex4-1}

\usepackage[utf8]{inputenc}

\usepackage{diagbox}

\usepackage{mathtools}
\usepackage{amsfonts}
\usepackage{mathrsfs}
\usepackage{bbm}
\usepackage{slashed}
\usepackage{braket}
\usepackage{titlesec}

\usepackage{graphicx}
\usepackage{color}
\usepackage{array}

\usepackage{blindtext}
\usepackage{placeins}
\usepackage{booktabs}
\usepackage{makecell}
\usepackage{epstopdf}
\usepackage[caption=false]{subfig}

\usepackage{xspace}
\usepackage{siunitx}
\usepackage{hyperref}
\usepackage[nameinlink]{cleveref}
\usepackage{appendix}

\usepackage{xifthen}
\usepackage{xcolor}
\hypersetup{
	colorlinks,
	linkcolor={blue!75!black},
	citecolor={blue!75!black},
	urlcolor={blue!75!black}
}

\setkeys{Gin}{width=0.48\textwidth}

\graphicspath{{figures/}}

\def\Eq#1{Eq.~\labelcref{#1}}

\usepackage{multirow}
\usepackage{colortbl}
\definecolor{kugray5}{RGB}{224,224,224}
\newcommand{\PreserveBackslash}[1]{\let\temp=\\#1\let\\=\temp}
\newcolumntype{C}[1]{>{\PreserveBackslash\centering}p{#1}}
\newcolumntype{R}[1]{>{\PreserveBackslash\raggedleft}p{#1}}
\newcolumntype{L}[1]{>{\PreserveBackslash\raggedright}p{#1}}

\newcommand{\gettitle}{Pion Distribution Amplitudes from Functional QCD}

\newcommand{\getHeidelbergAffiliation}{\affiliation{Institut f{\"u}r Theoretische Physik, Universit{\"a}t Heidelberg, Philosophenweg 16, 69120 Heidelberg, Germany}}

\newcommand{\getEMMIAffiliation}{\affiliation{ExtreMe Matter Institute EMMI, GSI, Planckstr. 1, D-64291 Darmstadt, Germany}}

\newcommand{\getDalianAFfiliation}{\affiliation{School of Physics, Dalian University of Technology, Dalian, 116024, P.R. China}}

\newcommand{\getNankaiAFfiliation}{\affiliation{School of Physics, Nankai University, Tianjin, 300071, P.R. China}}

\newcommand{\getUTokyoAFfiliation}{\affiliation{Institute for Physics of Intelligence, Graduate School of Science, The University of Tokyo, Bunkyo-ku, Tokyo 113-0033, Japan}}

\newcommand{\getRIKENAFfiliation}{\affiliation{RIKEN Center for Interdisciplinary Theoretical and Mathematical Sciences (iTHEMS), Wako, Saitama 351-0198, Japan}}

\hypersetup{
	pdftitle={\gettitle},
	pdfauthor={Chuang, Pawlowski},
	pdfkeywords={functional methods}
		{Distribution Amplitudes}
		{functional renormalisation group},
	bookmarksopen=true,
	bookmarksopenlevel=2,
	bookmarksnumbered=true
}

\begin{document}

\title{\gettitle}

\author{Lei Chang}
\getNankaiAFfiliation

\author{Wei-jie Fu}
\getDalianAFfiliation

\author{Chuang Huang}
\email{huang@thphys.uni-heidelberg.de}
\getHeidelbergAffiliation

\author{Jan M. Pawlowski}
\getHeidelbergAffiliation\getEMMIAffiliation

\author{Yang-yang Tan}
\getUTokyoAFfiliation\getRIKENAFfiliation


\begin{abstract}

We present the first functional QCD calculation of the pion distribution amplitude (DA) using the large-momentum effective theory within the functional renormalisation group (fRG) approach. With only the strong coupling and current quark masses as inputs, we compute the quasi-DA from first-principles QCD correlation functions. By pushing the pion momentum up to $P_z = 4.5\ \mathrm{GeV}$, the quasi-DA becomes fully saturated, rendering the extrapolation errors to the light-cone limit negligible. The resulting second-order moment $\langle \xi^2 \rangle_\pi = 0.267$ is significantly smaller than existing lattice-LaMET determinations and lies in a range consistent with other nonperturbative approaches.

\end{abstract}

\maketitle

\emph{Introduction.--}
The pion light-front parton distribution amplitude (PDA) describes how the pion's longitudinal momentum is shared between its valence quark and antiquark. As the pseudo-Goldstone boson of dynamical chiral symmetry breaking (DCSB)—the mechanism responsible for most of the visible mass in the universe—the pion's PDA provides a direct image of DCSB in a light-front wave function. Despite decades of study, the shape of the pion PDA remains controversial: different nonperturbative approaches yield conflicting results, with the second-order moment $\langle\xi^2\rangle_\pi$ ranging from $0.23$ to $0.30$. Resolving this tension is essential for understanding how DCSB manifests in hadron structure.

In this work, we present the first computation of the pion PDA within the functional QCD framework developed in~\cite{Fu:2022uow, Fu:2024ysj, Fu:2025hcm}. Unlike previous functional calculations that focused on integrated quantities such as meson masses, the PDA—as an $x$-dependent distribution—provides a more stringent test of the framework. Importantly, our approach contains no phenomenological parameters: only the strong coupling and current quark masses are fixed by $m_\pi/f_\pi$ and $m_K/f_\pi$, and all correlation functions are obtained self-consistently via the functional renormalisation group (fRG).

To access the full $x$-dependence without model-dependent moment reconstructions, we employ the large-momentum effective theory (LaMET)~\cite{Ji:2013dva, Ji:2020ect, Ji:2017rah}, which connects the light-cone PDA to a Euclidean quasi-PDA computed at large longitudinal momentum $P_z$. While lattice QCD has implemented LaMET~\cite{Zhang:2017bzy, LatticeParton:2022zqc}, the accessible $P_z$ is currently limited to $\sim 2.15$ GeV, where the quasi-PDA has not yet been saturated. We overcome this limitation by reaching $P_z = 4.5$ GeV—more than twice that of current lattice calculations—where the quasi-PDA is fully saturated, rendering the infinite-momentum extrapolation errors negligible.

We shortly summarise our main findings before discussing them in detail below. The resulting light-cone PDA exhibits a broad, concave, unimodal shape without a double-humped structure. The second-order moment is $\langle\xi^2\rangle_\pi = 0.267$, significantly smaller than the lattice-LaMET value ($0.300$) and consistent with other nonperturbative determinations, including lattice OPE~\cite{Arthur:2010xf, RQCD:2019osh}, QCD sum rules~\cite{Ball:2007zt, Zhong:2021epq}, and DSE/BSE calculations~\cite{Chang:2013pq, Roberts:2021nhw}. This work demonstrates that the functional QCD framework can provide first-principles, parameter-free predictions for parton distributions, opening a new avenue for studying hadron structure.

\begin{figure}[t]
\includegraphics[width=0.45\textwidth]{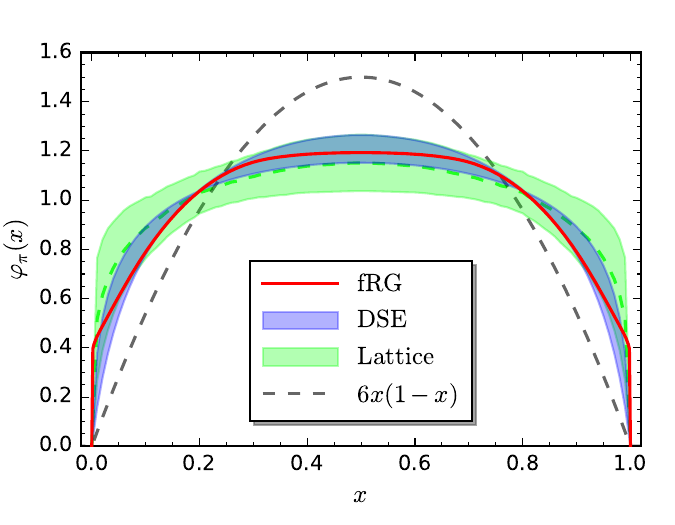}
\caption{Light-front PDA of pion as a function of the momentum fraction $x$. It is obtained by extrapolating quasi-light-front PDAs computed up to the longitudinal momentum $P_z=4.5$ GeV.  Our result is compared with the  results from lattice QCD \cite{LatticeParton:2022zqc} and from  DSE-BSE computations \cite{Chang:2025lrc}. We also show the asymptotic result $6x(1-x)$. }
\label{fig:PDA-final}
\end{figure}

\vspace{1ex}
\emph{Main result.--}
In \Cref{fig:PDA-final} we show the PDA obtained with LaMET, using QCD correlation functions obtained from the functional QCD approach~\cite{Fu:2025hcm}. Our result is  compared with those from lattice QCD~\cite{Arthur:2010xf, Braun:2015axa, Bali:2017ude, Zhang:2017bzy, RQCD:2019osh,  Loffler:2021afv, LatticeParton:2022zqc} and the functional DSE-BSE approach~\cite{Chang:2013pq, Shi:2018zqd, Cui:2020tdf, Raya:2021zrz, Roberts:2021nhw,  Xu:2025hjf, Chang:2025lrc}.

%
%

The quasi-light-front PDA calculations in this work are performed up to a maximal longitudinal momentum of $P_z=4.5$ GeV, in comparison to $P_{z}=2.15$ GeV in the lattice \cite{LatticeParton:2022zqc}. This allows us to access the interesting question of the saturation of the quasi-PDA at large $P_z$, and reduces significantly the errors in extracting the light-cone PDA from the quasi-PDA. Specifically, no sizable residual effect is observed. 

Furthermore, the quark two-point function and pion BSE inputs in this work are obtained from the functional QCD approach in \cite{Fu:2025hcm} without phenomenological parameters or model assumptions: only the fundamental parameters of (isospin-symmetric) QCD are fixed with the physical pion and kaon masses $m_\pi/f_\pi$ and $m_K/f_\pi$. This is achieved as the approximation in \cite{Fu:2025hcm} accommodates in particular the complete set of dominant and subdominant four-quark operators and quark-gluon tensor structures relevant for the full dynamics of spontaneous chiral symmetry breaking in QCD. This constitutes a significant step forward in bound-state studies in functional QCD. Our PDA shows a flat behaviour in the intermediate-$x$ region, indicating a stronger impact of DCSB in this regime. Compared to the lattice-QCD result, our PDA is slightly narrower, which indicates that the effect of DCSB is more pronounced at larger $P_z$.

%
\begin{figure}[t]
\includegraphics[width=0.5\textwidth]{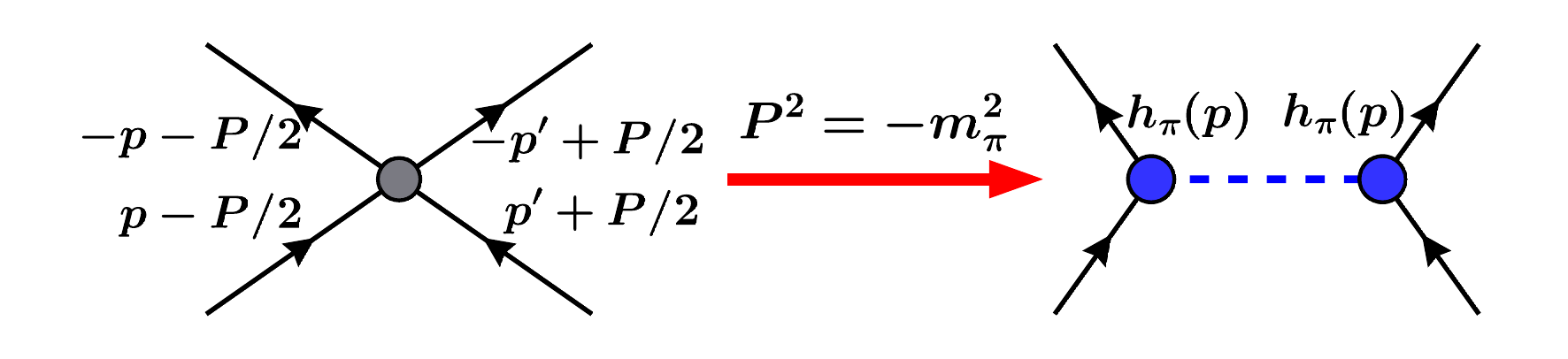}
\caption{Schematic illustration of the four-quark vertex and its reduction to the (on-shell) pion exchange for $P^2\to -m_\pi^2$. The blue dashed line denotes the meson propagator, and the blue circles represent the BS amplitudes. The momenta are defined in \Eq{eq:Pp-p1234} and \Eq{eq:AngularFix}. All the momenta are incoming towards the vertices, and the arrows indicate the direction of fermion flows.}
\label{fig:resonance}
\end{figure}
%

\vspace{1ex}
\emph{2+1 flavour functional QCD.--}
The computation of the PDAs with LaMET is based on the quasi-light-front wave function (quasi-LFWF) \labelcref{eq:Q-LFWF+BS-Amplitude}. The quasi-LFWF is constructed from the quark propagator and the resonant four-quark correlation functions of pion channel in QCD. Here, they are provided within the fRG approach to QCD. Specifically, we use the setup \cite{Fu:2025hcm} that also builds on the advances in \cite{Fu:2022uow, Fu:2024ysj, Fu:2025hcm}. Apart from the many tests and systematic error checks in these works, they are embedded in the fRG approach to QCD, systematically developed in \cite{Mitter:2014wpa, Braun:2014ata, Rennecke:2015eba, Cyrol:2016tym, Cyrol:2017ewj, Corell:2018yil, Fu:2019hdw, Ihssen:2024miv, Fu:2025hcm, Pawlowski:2025jpg, Fu:2026qnl}, mostly within the fQCD collaboration \cite{fQCD}, for recent reviews see \cite{Dupuis:2020fhh, Fu:2022gou, Rennecke:2025bcw, Fischer:2026uni, Fischer:2026vkc}. Functional approaches have also provided in-depth studies of DCSB and meson physics in the vacuum, both at the structural and quantitative level, see e.g.~\cite{Gao:2024gdj, Miramontes:2025imd, Ferreira:2025wpu, Miramontes:2026san, Ferreira:2026gbe}. These works provide a comprehensive error and self-consistency analysis for functional QCD at finite temperature and baryon densities, and in particular for vacuum QCD.  More details are provided in the supplement. We work in the isospin-symmetric limit with the light and strange current quark masses $m_l,m_s$ with $m_u=m_d=m_l$. These parameters are fixed with the ratio of the pion and kaon masses with the pion decay constant, $m_\pi / f_\pi$ and $m_K/f_\pi$, for more details see \cite{Fu:2025hcm} and the supplement. 

The setup of 2+1 flavour functional QCD within the fRG approach is recapitulated in the supplement, where the full quantum effective action is comprised of the glue sector, the quark-gluon interactions of classical and non-classical tensor structures, and the multiple quark scatterings in the regime of low energy. Note that the convergence of truncations in the expansion of correlation functions was observed and demonstrated with the increase of $n$-point functions in the previous studies, see the references above as well as, e.g., \cite{Braun:2025gvq, Eichmann:2026ttr} for more details. Rapid convergence requires that, in particular the light resonances emerging in the low-energy region are taken into account properly. This concerns in particular the pion and $\sigma$-modes, but all light resonances that are relevant for the off-shell dynamics of low energy QCD emerge from the four-quark vertices. In conclusion, a comprehensive resolution of the four-quark scattering vertex is chiefly important for studies of hadron structure.

The four-quark scattering vertex is expanded with a Fierz-complete tensor basis $\{{\cal T}^\alpha(p_1,p_2,p_3,p_4)\}$, ordered in terms of their momentum dimensions. In \cite{Fu:2022uow, Fu:2024ysj, Fu:2025hcm} all momentum-independent tensor structures and their crossing-symmetric partners were considered. Their dressings are factorised into a product of $Z^{1/2}_l(p_i)$ for each quark leg with momenta $p_i$ and the RG-invariant dressings $\lambda^\alpha(p_1,p_2,p_3,p_4)$ with $p_4=-(p_1+p_2+p_3)$ can be parametrised in three radial momenta and three angles. For their dependence we take the Mandelstam variables $s$, $t$, and $u$,  
\begin{align}
    s=(p+p')^2,\quad t=P^{2},\quad u=(p-p')^2\,.
    \label{eq:stu-general}
\end{align}
with 
\begin{align}
    P=-(p_{1}+p_{2}),\quad p=\frac{p_{2}-p_{1}}{2},\quad p'=\frac{p_{4}-p_{3}}{2}\,,
    \label{eq:Pp-p1234}
\end{align}
It was checked in \cite{Fu:2022uow, Fu:2024ysj, Fu:2025hcm}, that the magnitude and angle dependence of four-quark dressings are well captured with the reduction of dependence on the three Mandelstam variables, with the configuration $p=|p|(1,0,0,0)$ and $p'=|p|(\cos \theta, \sin \theta, 0, 0)$. This leads us to 
\begin{align}
    \lambda_{\alpha}(p_{1},p_{2},p_{3},p_{4})\approx  \lambda_{\alpha}(s,t,u)\,, 
\label{eq:lambda-stu}
\end{align}
with $\alpha\in \{\sigma,\pi,\kappa,K\}$. This approximation is quantitatively reliable and the resulting errors were studied in detail in \cite{Fu:2024ysj, Fu:2025hcm}, which were found to be less than 1.5\% combined with the configuration in \Eq{eq:AngularFix}.  Heuristically, this originates in the fact that in the infrared momentum regime, where these scattering processes give sizable contributions to the off-shell dynamics of QCD, they are dominated by the mesonic resonant momentum channels. This is illustrated in \Cref{fig:resonance} where the  $t$-channel of the four-quark vertex is shown. For example, for $t=-m_\pi^2$ with the pole mass $m_\pi$ of the pion, the pseudoscalar channel diverges and simply reduces to the product of the Bethe-Salpeter (BS) wave function, its conjugate and the pion on-shell propagator $1/(t+m_\pi^2)$. Finally, for the computation of \Eq{eq:Q-LFWF+BS-Amplitude} we choose the symmetric configuration,  
\begin{align}
    p=-p'=|p|\left(\cos\theta,\sin\theta,0,0\right)\,,\label{eq:AngularFix}
\end{align}
and read off $\lambda_{\alpha}(P,p,\cos\theta)$ from the integrated flow. 

In summary, this allows us to extract the BS amplitude, which is simply given by the residue of the dressings at the bound-state pole. For the pion meson, its RG-invariant BS amplitude $h_\pi$ is given by
\begin{align}
    \hspace{-0.2cm}h_{\pi}(p,\cos\theta)= \lim_{P^2\to -m_{\pi}^2} \sqrt{\lambda_{\pi}(P^2,p,\cos\theta)(P^2+m_{\pi}^2)} \,.
\end{align}
%

%
\begin{figure}[t]
\includegraphics[width=0.45\textwidth]{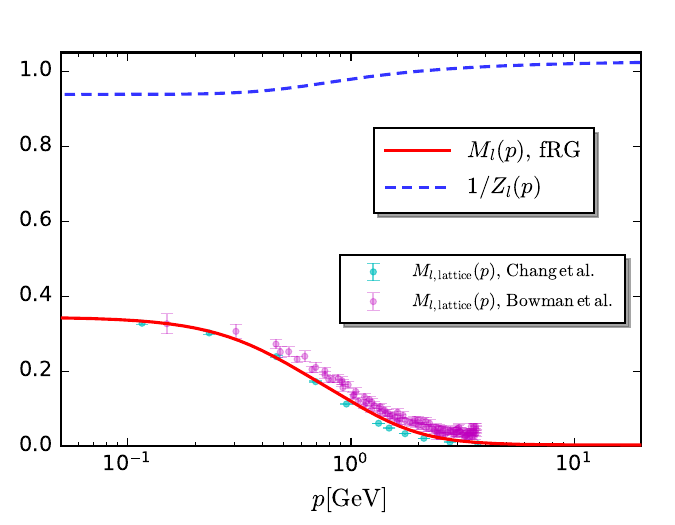}
\caption{Light quark mass function $M_{l}(p)$ (solid line, GeV) and the quark wave function $Z_{l}(p)$ at renormalisation point $\mu=2$ GeV (dashed line). The lattice data are taken from \cite{Chang:2021vvx} (cyan points) and \cite{Bowman:2005vx} (purple points).}\label{fig:Quark}
\end{figure}
%

\begin{figure}[t]
\includegraphics[width=0.45\textwidth]{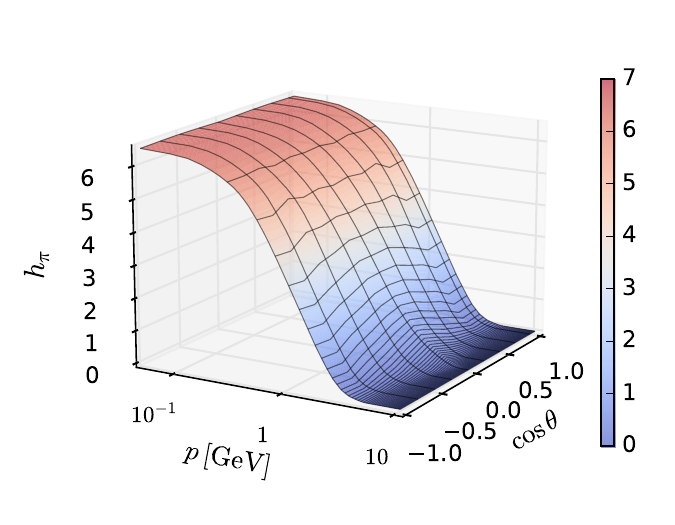}\hspace{0.5cm}
\caption{Bethe-Salpeter amplitude of the pion as a function of the magnitude of the quark momentum and the angle between the quark and meson momenta.}\label{fig:BS-amplitude}
\end{figure}

With these preparations, we proceed with the QCD input required for the computation of the quasi-LFWF, cf. \Eq{eq:Q-LFWF+BS-Amplitude}: In \Cref{fig:Quark} we show the light quark wave function and mass function, and in \Cref{fig:BS-amplitude} we show the pion Bethe-Salpeter amplitude. The RG-invariant light quark mass function agrees quantitatively with the lattice QCD results in \cite{Chang:2021vvx, Bowman:2005vx}. The inverse of the wave function $1/Z_l$ exhibits a slight decrease in the infrared region, see also \cite{Mitter:2014wpa, Williams:2014iea, Williams:2015cvx, Aguilar:2016lbe, Cyrol:2017ewj, Gao:2021wun, Ihssen:2024miv, Aguilar:2024ciu}. For the pion BS amplitude, it is worth emphasising that we observe a mild angular dependence. As the relative angle deviates from $\cos\theta=0$, the momentum dependence of the BS amplitude becomes slightly steeper.

%
%

\vspace{1ex}
\emph{Pion distribution amplitude on the quasi-LF and LF.--}
With the correlation functions obtained from the above first-principles functional QCD computation and method developed in \cite{Zhang:2025ofc,Cui:2026bod}, we proceed to the calculation of the pion PDA within both the quasi-light-front (LF) and LF frameworks.

In the quasi-light-front framework, we introduce the longitudinal momentum $P_{z}$ in the on-shell momentum of the pion, which is given by
\begin{align}
    P_{\mu} = (\mathrm{i}E_{\pi}, P_{z},0,0)\,,\quad E_{\pi} = \sqrt{P_{z}^2+m_{\pi}^2}\,,
\end{align}
where $P^2=-m_{\pi}^2$ is Lorentz invariant. Then the quasi-light-front wave function is defined as 
\begin{subequations}
    \label{eq:Q-LFWF+BS-Amplitude}
\begin{align}
    &\psi_\pi (x,P_{z},p_{\perp}) \nonumber\\[2ex]&\hspace{-.2cm}= \frac{1}{f_{\pi}}\int \frac{dp_{0}dp_{3}}{\pi}\delta(\tilde n\cdot p_+-x\tilde n\cdot P)\gamma_5 \, \tilde n\cdot \gamma\, \chi_\pi(p;P)\,,
    \label{eq:Q-LFWF}
\end{align}
where $x$ denotes the longitudinal momentum fraction of the valence quark in pion, and $f_{\pi}$ is the pion decay constant. $\tilde n = (0,1,0,0)$ denotes the longitudinal direction in the Euclidean space. The unamputated Bethe-Salpeter amplitude $\chi_\pi$ in \labelcref{eq:Q-LFWF} is obtained from the BS-amplitude 
$\Gamma_\pi$ by attaching the light quark propagators $G_{l}$, i.e.,
\begin{align}
    \chi_\pi(p;P) = G_{l}(p_{+})\Gamma_{\pi}(p;P)G_{l}(p_{-})\,,
    \label{eq:BS-Amplitude}
\end{align}
with
\begin{align}
    \Gamma_{\pi}(p;P)=\mathrm{i}\gamma_5\,Z_{l}^{1/2}(p_{+})Z_{l}^{1/2}(p_{-}) h_{\pi}(p;P) \,.
\end{align}
\end{subequations}
Here, $p_{\pm} = p \pm P/2$ denotes the momenta of two quark propagators in the BS amplitude. The light quark propagator is built up from the dressing of the Dirac part, the wave function $1/Z_{l}(p)$, and the mass function $M_{l}(p)$. The pion BS amplitude $h_{\pi}$ depends on the pion momentum $P$ and the relative momentum $p$. All these different parts of $\chi_\pi(p;P)$ are shown in  \Cref{fig:Quark} and \Cref{fig:BS-amplitude}. All parts except for $Z_l$ are RG-invariant and an RG-transformation of the latter leads to a global rescaling of the unamputated BS-amplitude $\chi_\pi$. Our explicit results are achieved with an RG-point $\mu=2$\,GeV with $Z_l(\mu)=1$.   

%
\begin{figure}[t]
\includegraphics[width=0.45\textwidth]{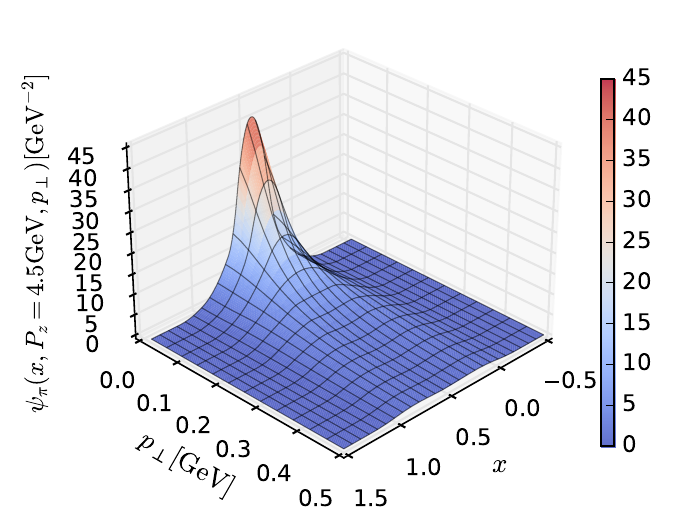}
\caption{3D plot of the pion quasi-LFWF as a function of the momentum fraction $x$ and transverse momentum $p_{\perp}$ with the longitudinal momentum $P_z=4.5$ GeV.}\label{fig:wave-function}
\end{figure}
%

To compute the quasi-LFWF in \Eq{eq:Q-LFWF+BS-Amplitude}, one also needs the information about the analytic properties of the quark propagator and BS amplitude on the complex plane of $p_0$, see \cite{Eichmann:2021vnj,dePaula:2022pcb}. We use the Taylor expansions to include the relevant information as done in \cite{Zhang:2025ofc}, which in turn enlarges the accessible range of longitudinal momentum $P_{z}$ substantially and improves the convergence of the PDA by extrapolating $P_{z} \to \infty$. It is found that the expansion up to the fourth order of $p_0$ is sufficient to obtain convergent results, and more relevant discussions are presented in the supplement. In \Cref{fig:wave-function}, we show the unnormalised pion quasi-LFWF $\psi_\pi(x,p_{\perp})$ at $P_z=4.5$ GeV. Since the pion consists of two light quarks, the quasi-LFWF is symmetric with respect to $x=0.5$. There is a peak structure in the intermediate region around $x=0.5$, which descends rapidly with the increasing transverse momentum $p_{\perp}$.

%
\begin{table*}[t]
    \begin{center}
    \centering
    \begin{tabular}{lccc}
      \hline\hline & & &   \\[-2ex]   
      Method & $\langle\xi^2\rangle_\pi$ & $\langle\xi^4\rangle_\pi$ & $\langle\xi^6\rangle_\pi$   \\[1ex]
      \hline & & &   \\[-2ex]
      functional LaMET (This Work) & $0.267$ & $0.139$ & $0.088$   \\[1ex]
      Lattice LaMET (LPC)\cite{LatticeParton:2022zqc} & $0.300(41)$ & - & -   \\[1ex]
      Lattice OPE (RQCD)\cite{RQCD:2019osh} & $0.234^{+6}_{-6}(4)(4)(2)$ & - & -   \\[1ex]
      Lattice OPE (RBC and UKQCD)\cite{Arthur:2010xf} & $0.28(1)(2)$ & - & -   \\[1ex]
      Sum Rule\cite{Ball:2007zt, Zhong:2021epq}& $0.271(13)$ & $0.138(10)$ & $0.087(6)$   \\[1ex]
      DSE/BSE [RL, DB]\cite{Chang:2025lrc} & $0.280,\,0.247$ & $0.149,\,0.121$ & $0.097,\,0.073$   \\[1ex]
      \hline\hline
    \end{tabular}
    \caption{Moments of the valence-quark parton distributions of the pion up to the sixth order calculated in this work. The results are compared with those of lattice QCD based on the LaMET \cite{LatticeParton:2022zqc} and the DSE/BSE \cite{Chang:2025lrc}. Moreover, we also present the results of lattice QCD based on the Operator Product Expansion (OPE) \cite{Arthur:2010xf, RQCD:2019osh} and QCD sum rules \cite{Ball:2007zt, Zhong:2021epq}.}
    \label{tab:pion-moments}
    \end{center}\vspace{-0.5cm}
\end{table*}
%

\begin{figure}[t]
\includegraphics[width=0.45\textwidth]{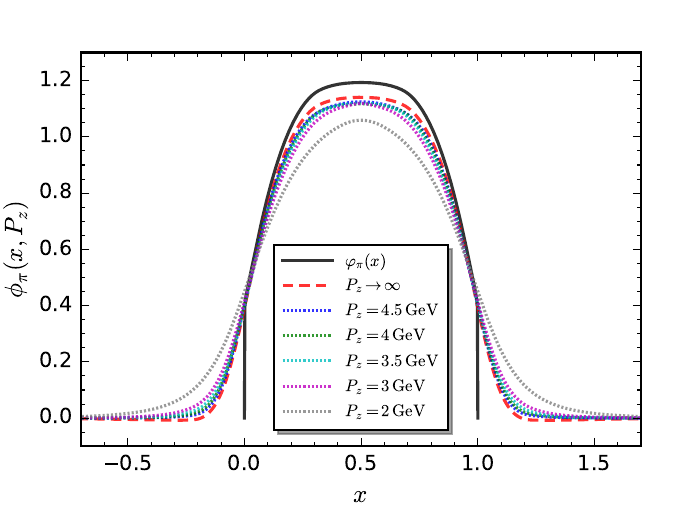}
\caption{PDA and quasi-PDA of pions as functions of the momentum fraction $x$. The quasi-PDAs at $P_{z}=2$, 3, 3.5, 4, 4.5 GeV are shown by the dotted lines of different colours. The PDA obtained with only the extrapolation of $P_z \to \infty$ is shown by the red dashed line, and the PDA with both extrapolations in $P_z \to \infty$ and the endpoint regions of $x$ is shown by the black solid line.}
\label{fig:quasi-PDA}
\end{figure}

The quasi-PDA is defined as the integral of the quasi-LFWF over the transverse momentum, viz.,
\begin{align}
    \phi_\pi(x, P_z) = \frac{1}{16 \pi^3}\int d^2 p_{\perp} \psi_\pi (x,P_{z},p_{\perp})\,,\label{eq:qPDA}
\end{align}
from which, one is able to obtain the pion PDA $\varphi_{\pi}(x)$ on the light cone by extrapolating \Eq{eq:qPDA} in the large longitudinal momentum limit, i.e.,
\begin{align}
    \varphi_{\pi}(x)=\phi_{\pi}(x,P_{z}\to\infty)\,.
\end{align}
In \Cref{fig:quasi-PDA}, we present the normalised pion PDA $\varphi_\pi(x)$ and the quasi-PDA $\phi_\pi(x,P_z)$. The coloured dotted curves show the quasi-PDA at different $P_{z}$, and they converge as $P_{z}$ increases. It is found that as $P_{z}\gtrsim 3.5$ GeV, the quasi-PDA is already saturated, indicating the extrapolation to large $P_{z}$ limit is stable and convergent. The red dashed line denotes the extrapolated PDA in the limit $P_z \to \infty$, which lies very close to the large-$P_{z}$ quasi-PDA. As expected, the extrapolation shows a linear dependence on the $1/P_{z}^{2}$ term. Notably, due to the limitations of the Euclidean correlation functions inputs and the LaMET approach near the endpoints $x=0$ and $x=1$, we also perform extrapolations in the endpoint regions of $x$ and obtain the final PDA shown by the black solid line. Varying the endpoint fitting range changes the result only marginally, indicating that the endpoint extrapolation only gives rise to a very small uncertainty.

%
%

In \Cref{fig:PDA-final} we have compared our results for the pion PDA with those from lattice QCD based on LaMET and DSE calculations. 

The moments of the PDA are defined as
\begin{align}
    \langle\xi^n\rangle_\pi = \int dx\, (2x-1)^n \varphi_\pi(x)\,. 
\end{align}
They can be used for a more detailed comparison of the results from different approaches. In \Cref{tab:pion-moments}, we present the second, fourth, and sixth-order moments of the PDA. The second-order moment of functional-LaMET PDA is smaller than the lattice-LaMET result and closer to those from other approaches, such as the lattice OPE \cite{Arthur:2010xf, RQCD:2019osh}, QCD sum rules \cite{Ball:2007zt, Zhong:2021epq}, and DSE/BSE \cite{Chang:2013pq, Roberts:2021nhw, Chang:2025lrc}. The higher-order moments of the PDA from functional LaMET are also consistent with the sum rules and DSE/BSE. In particular, compared with the lattice-LaMET determination, our second-order moment is significantly smaller and falls into a range comparable with other approaches. This is likely because the access to larger $P_z$ provides for a more complete inclusion of the relevant QCD dynamics at higher scales, and hence a more faithful incorporation of DCSB effects, which leads to a narrower PDA.

\vspace{1ex}
\emph{Conclusions.--}
In this work, we have directly computed the pion quasi-PDA at finite longitudinal momenta from first-principles functional QCD. Specifically, we have used the fRG approach to QCD and extrapolated the light-cone PDA in the limit of large longitudinal momentum based on the large-momentum effective theory (LaMET). In our calculations, the maximal longitudinal momentum has been extended up to $P_z \sim 4.5\,\mathrm{GeV}$. This is achieved with the techniques developed in \cite{Zhang:2025ofc}, e.g., the deformed integration contour, the analytic information of the quark propagator and pion Bethe-Salpeter amplitude in the complex plane of $p_0$ encoded through Taylor expansions. Importantly, we directly observe the saturation of the quasi-PDA with the increase of $P_z$. This direct access is achieved due to the significant extension of the regime of available longitudinal momenta in comparison to the lattice calculation with the maximal momentum $P_z \sim 2.15\,\mathrm{GeV}$ \cite{LatticeParton:2022zqc}. This saturation ensures the accuracy of the extrapolation of $P_z\to\infty$ and reduces the associated errors to a negligible level. We have compared the pion light-cone PDA obtained from our functional QCD based on the LaMET with the lattice-QCD LaMET result as well as other approaches. We find that the second-order moment of pion PDA from the functional-QCD LaMET is significantly smaller than the lattice-QCD LaMET result. Indeed, it lies in a range comparable to other approaches which resolves a longstanding conflict between the lattice-QCD LaMET and other nonperturbative approaches concerning the variation of pion parton distributions.

\vspace{1ex}
\emph{Acknowledgements.--}
We thank Gernot Eichmann, Fei Gao, Jun Hua, Joannis Papavassiliou, Bernd-Jochen Schaefer and Jonas Wessely for discussions. This work is funded by the Deutsche Forschungsgemeinschaft (DFG, German Research Foundation) under Germany’s Excellence Strategy EXC 2181/1 - 390900948 (the Heidelberg STRUCTURES Excellence Cluster), the Collaborative Research Centre SFB 1225 - 273811115 (ISOQUANT), the National Natural Science Foundation of China under Grant No.\ 12447102, the JSPS KAKENHI Grant No. 25H01560, and JST-BOOST Grant No. JPMJBY24H9. It is also supported by EMMI. JMP acknowledges support by the Chinese Academy of Sciences President's International Fellowship Initiative Grant No.~2024PG0023.

\hfill
\bibliography{ref-lib}

\newpage

\renewcommand{\thesubsection}{{S.\arabic{subsection}}}
\setcounter{section}{0}
\titleformat*{\section}{\centering \Large \bfseries}

\onecolumngrid

\section*{Supplemental Material}

We provide supplemental material about the fRG approach to QCD, \Cref{sec:2+1QCD}, the contour deformation used for the computation of the quasi-PDA, \Cref{sec:integrate-contour}, the Taylor expansion of the BSE-amplitude and the quark mass, \Cref{sec:expansion}, and the extrapolation of the large longitudinal momentum results, \Cref{sec:extrapolation}.


\subsection{2+1 flavour functional QCD}
\label{sec:2+1QCD}

In this Appendix, we recapitulate the fRG approach to 2+1 flavour functional QCD developed in \cite{Fu:2025hcm}. This approach allows for a self-consistent of QCD correlation functions without any phenomenological parameters and external inputs. One can clearly identify three sectors: the pure glue, glue-matter interface, and the pure matter sectors. This separation into three sectors provides a simple comprehensive access to the overall systematic error of the approach. It can be built up from the systematic errors of the sectors and by considering the combined error propagation, for more details see \cite{Ihssen:2024miv, Fu:2025hcm, Fischer:2026uni}. 

We begin with discussion with the decomposition of the quantum effective action of the 2+1 flavour QCD into the three sectors 
\begin{align}
	\Gamma_{k}[\Phi]=\Gamma_{\mathrm{glue},k}[A,\bar{c},c]+\Gamma_{\textrm{inter},k}[A,\bar{q},q]+\Gamma_{4q,k}[\bar{q},q]\,.  \label{eq:QCD-action}
\end{align}
The fields $\Phi=(A,c,\bar{c},q,\bar{q})$ include the gluon, ghost, anti-ghost, quark and anti-quark, respectively. The subscript $k$ denotes the infrared cutoff, or the renormalisation group (RG) scale in the RG flows. 

In the present work we consider the following approximation of the three sectors in \Eq{eq:QCD-action}, 
\begin{align}
    \Gamma_{\mathrm{glue},k}[A,\bar{c},c]  &=\frac12 \int\limits_p \, A^a_\mu(p) \, \left[\Gamma_{AA}^{(2)}\Pi^{\perp}\right]^{ab}_{\mu\nu}(p) \, A^a_\nu(-p) + 	\int\limits_p \bar c^{\,a}(p)\left[\Gamma_{c\bar c}^{(2)}\right]^{ab}(p)\, c^b(-p)\nonumber \\[2ex]
    &\quad+ \int\limits_{p_1,p_2} \left[\Gamma_{A\bar c c}^{(3)}\right]^{a_1 a_2 a_3}_{\mu}(p_1,p_2,p_3)\, \bar c^{a_2} (p_2) c^{a_1}(p_1) A^{a_3}_\mu(p_3) \nonumber \\[2ex]  
    &\quad + \frac{1}{3!} \int\limits_{p_1,p_2}  \left[\Gamma_{A^3}^{(3)}\right]^{a_1 a_2 a_3}_{\mu_1\mu_2\mu_3}(p_1,p_2,p_3) \prod_{i=1}^3 A^{a_i}_{\mu_i}(p_i)   \nonumber \\[2ex]  
    &\quad+  \frac{1}{4!}\int\limits_{p_1,p_2,p_3}  \left[\Gamma_{A^4}^{(4)}\right]^{a_1 a_2 a_3 a_4}_{\mu_1\mu_2\mu_3\mu_4}(p_1,p_2,p_3,p_4) \prod_{i=1}^4 A^{a_i}_{\mu_i}(p_i) \,,\label{}\\[4ex]
    \Gamma_{\textrm{inter},k}[A,\bar{q},q]  &=\int\limits_p \bar q(-p)\left[\Gamma_{\bar q q}^{(2)}\right](p)\, q(p)+ \int\limits_{p_1,p_2} \left[\Gamma_{A\bar q q}^{(3)}\right]^{a}_{\mu}(p_1,p_2,p_3)\,\bar q (p_2) A^{a}_\mu(p_1) q(p_3)\nonumber\\[2ex] 
    &\quad+\int\limits_{p_1,p_2,p_3} \left[\Gamma_{AA\bar q q}^{(4)}\right]^{ab}_{\mu_1 \mu_2}(p_1,p_2,p_3,p_4)\,\bar q (p_3) A^{a}_{\mu_1}(p_1)A^{b}_{\mu_2}(p_2) q(p_4) \,,\label{}\\[4ex]
    \Gamma_{4q,k}[\bar{q},q]  &=\int\limits_{p_1,p_2,p_3} \left[\Gamma_{4q}^{(4)}\right]^{(\alpha)}_{ijkm}(p_1,p_2,p_3,p_4)\bar q_{i} (p_1)  q_{j} (p_2) \bar q_k (p_3) q_m (p_4)\,,\label{}
\end{align}
respectively, with
\begin{align}
    \int_{p}\equiv\int \frac{d^{4}p}{(2\pi)^{4}}\,.
\end{align}
The two-point functions for the gluon and ghost fields read
\begin{align}
    \left[\Gamma_{AA}^{(2)}\Pi^{\perp}\right]^{ab}_{\mu\nu}(p) = Z_A(p)p^2 \delta^{ab} \Pi^{\perp}_{\mu\nu}(p) \,,\qquad\left[\Gamma_{c\bar c}^{(2)}\right]^{ab}(p) = Z_c(p)p^2 \delta^{ab} \,,
\end{align}
where $Z_A(p)$ and $Z_c(p)$ are the wave functions for the gluon and ghost fields, respectively. Here, 
\begin{align}
    \Pi^{\perp}_{\mu\nu}(p)=\delta_{\mu\nu} - \frac{p_\mu p_\nu}{p^2}\,,
\end{align}
denotes the transverse projection operator. For the quark two-point function, in 2+1 flavour QCD, we consider light quarks and strange quarks. Within the isospin symmetry the quark two-point function reads
\begin{align}
    \Gamma_{\bar q q}^{(2)}(p) = Z_{q}(p)\,\Big[\mathrm{i}\slashed{p}+ M_{q}(p)\Big] \,,
\end{align}
with
\begin{align}
    Z_{q}(p) &= \text{diag}(Z_l(p), Z_l(p), Z_s(p)) \,,\\[2ex]
    M_{q}(p)&=\text{diag}(M_l(p), M_l(p), M_s(p))\,,
\label{eq:ZqMq}
\end{align}
where $Z_l(p)$ and $Z_s(p)$ represent the wave functions and $M_l(p)$ and $M_s(p)$ the mass functions for light and strange quarks, respectively. Consequently, the quark propagators read
\begin{align}
    G_l(p)=\frac{1}{Z_{l}(p)}\frac{-\mathrm{i}\gamma \cdot p+M_l(p)}{p^2 +M_l^2(p)}\,, \quad
    G_s(p)=\frac{1}{Z_{s}(p)}\frac{-\mathrm{i}\gamma \cdot p+M_s(p)}{p^2 +M_s^2(p)}\,.\label{eq:Gq}
\end{align}
The quark mass function \labelcref{eq:ZqMq} and the four-quark vertex dressings  \labelcref{eq:4q-lambda} are RG-invariant, that is 
\begin{align} 
\mu\frac{d}{d\mu} M_q(p) = 0\,,\qquad \mu\frac{d}{d\mu} \lambda_\alpha(p)=0\,, 
\label{eq:RG-invariant}
\end{align} 
with the RG-scale $\mu$. 
In turn, the wave function $Z_l(p)$ scales with the RG-scale. The standard fRG renormalisation scheme is the MOM-type MOM$^2$ renormalisation scheme, which is discussed in detail in \cite{Gao:2021wun} and has been used in all first-principles  QCD studies with the fRG, see e.g.~\cite{Mitter:2014wpa, Cyrol:2017ewj, Ihssen:2024miv,Fu:2025hcm}. Compared with the conventional MOM scheme, this choice allows us to set the wave-function to $Z_{i}(p = \mu_\textrm{UV})=1$ at a large RG-scale $\mu_\textrm{UV}$. The conversion between $Z_i(p)$ in the MOM$^2$ scheme and other renormalisation schemes amounts to an overall multiplicative constant.  

From \Eq{eq:Q-LFWF+BS-Amplitude}, the quasi-DA and unamputated BS amplitude are not RG-invariant quantities. The quark wave-functions determine their renormalisation. In the present work, we choose the renormalisation condition
\begin{align}
    Z_{l}(p=\mu)=1\,,\label{eq:Zq-renor}
\end{align}
with a renormalisation point $\mu=2$ GeV. In this way, our results are directly comparable with those obtained from other nonperturbative approaches, see also \cite{Cui:2026bod}.

Next, we introduce the vertex functions with RG-invariant dressings $\lambda_{i}$. First, the ghost-gluon vertex is given by
\begin{align}
    \left[\Gamma_{A\bar c c}^{(3)}\right]^{a_1 a_2 a_3}_{\mu}(p_1,p_2,p_3)& = (2\pi)^4 \delta^{(4)}(p_1+p_2+p_3)\nonumber\\[2ex]
    &\quad\times\left[Z_{c}^{\frac{1}{2}}(p_1)Z_{c}^{\frac{1}{2}}(p_2)Z_{A}^{\frac{1}{2}}(p_3)\right]\lambda_{A \bar c c}(p_1,p_2) \left[ {\cal T}_{A\bar c c}(p_1,p_2)\right]^{a_1 a_2 a_3}_{\mu} \,,
\end{align}
where ${\cal T}_{A\bar c c}$ is the classical tensor structure of the ghost-gluon vertex, given by
\begin{align}
    \left[ {\cal T}_{A\bar c c}(p_1,p_2)\right]^{a_1 a_2 a_3}_{\mu} = \mathrm{i}f^{abc} (p_2)_\mu \,,
\end{align}
where $f^{abc}$ are the structure constants of the $SU(N_c)$ group. 

For the ghost-gluon vertex dressing, here we introduce the symmetric point approximation, namely
\begin{align}
    \lambda_{A \bar c c}(p_1,p_2)\approx \lambda_{A \bar c c}(\bar{p})\,,
\end{align}
where $\bar{p}$ is the average momentum, given by
\begin{align}
    \bar{p}^2 = \frac{p_1^2+p_2^2+(p_1 +p_2)^2}{3}\,.
\end{align}
Under the symmetric point approximation, the dressing can be simplified from a multi-dimensional momentum function to a single-dimensional momentum function depending on the average momentum. This approximation will be applied to all vertex functions except the four-quark vertex, namely
\begin{align}
    \lambda_{i}(p_{1},p_{2},\cdots,p_{n_i})\approx \lambda_{i}(\bar{p})\,,
\end{align}
where $\lambda_{i}$ is the dressing of the vertex with $i\in \{A \bar c c, A^3, A^4, A\bar q q, AA\bar q q\}$. For $n$-point functions, the specific momentum configuration is chosen as
\begin{align}
    p_i \cdot p_j = \frac{ n\delta_{ij} -1}{n-1} \bar p^2\,,\quad \bar p^2 = \frac{1}{n}\sum_{i=1}^n p_i^2\,.
\end{align}
and the vertex momenta satisfy momentum conservation,
\begin{align}
    p_n = -(p_1 + p_2 + \cdots + p_{n-1})\,.
\end{align}
Next, we specifically introduce the other vertex functions. First, the three-gluon vertex is given by
\begin{align}
    \left[\Gamma_{A^3}^{(3)}\right]^{a_1 a_2 a_3}_{\mu_1\mu_2\mu_3}(p_1,p_2,p_3)& = (2\pi)^4 \delta^{(4)}(p_1+p_2+p_3)\nonumber\\[2ex]
    &\quad\times\left[\prod_{i=1}^3 Z_{A}^{\frac{1}{2}}(p_i)\right]\lambda_{A^3}(\bar{p}) \left[ {\cal T}_{A^3}(p_1,p_2,p_3)\right]^{a_1 a_2 a_3}_{\mu_1\mu_2\mu_3} \,,
\end{align}
where ${\cal T}_{A^3}$ is the classical tensor structure of the three-gluon vertex, given by
\begin{align}
    \left[ {\cal T}_{A^3}(p_1,p_2,p_3)\right]^{a_1 a_2 a_3}_{\mu_1\mu_2\mu_3} &= -\mathrm{i}f^{a_1 a_2 a_3} \Bigl[\delta_{\mu_1 \mu_2} (p_1-p_2)_{\mu_3} + \delta_{\mu_2 \mu_3} (p_2-p_3)_{\mu_1} + \delta_{\mu_3 \mu_1} (p_3-p_1)_{\mu_2} \Bigr] \,.
\end{align}
For the four-gluon vertex, it is given by
\begin{align}
    \left[\Gamma_{A^4}^{(4)}\right]^{a_1 a_2 a_3 a_4}_{\mu_1\mu_2\mu_3\mu_4}(p_1,p_2,p_3,p_4) &= (2\pi)^4 \delta^{(4)}(p_1+p_2+p_3+p_4)\nonumber\\[2ex]
    &\quad\times\left[\prod_{i=1}^4 Z_{A}^{\frac{1}{2}}(p_i)\right]\lambda_{A^4}(\bar{p}) \left[ {\cal T}_{A^4}(p_1,p_2,p_3,p_4)\right]^{a_1 a_2 a_3 a_4}_{\mu_1\mu_2\mu_3\mu_4} \,,
\end{align}
where ${\cal T}_{A^4}$ is the classical tensor structure of the four-gluon vertex, given by
\begin{align}
    \left[ {\cal T}_{A^4}(p_1,p_2,p_3,p_4)\right]^{a_1 a_2 a_3 a_4}_{\mu_1\mu_2\mu_3\mu_4} &= \left[f^{b a_1 a_2}f^{b a_3 a_4}(\delta_{\mu_1\mu_3}\delta_{\mu_2\mu_4} - \delta_{\mu_1\mu_4}\delta_{\mu_2\mu_3})\right.\nonumber   \\[2ex] 
    &\quad+f^{b a_1 a_3}f^{b a_2 a_4}(\delta_{\mu_1\mu_2}\delta_{\mu_3\mu_4} - \delta_{\mu_1\mu_4}\delta_{\mu_2\mu_3})
    \nonumber   \\[2ex] 
    & \quad \left. +f^{b a_1 a_4}f^{b a_2 a_3}(\delta_{\mu_1\mu_2}\delta_{\mu_3\mu_4} -\delta_{\mu_1\mu_3}\delta_{\mu_2\mu_4})\right] \,.
\end{align}
For the light quark-gluon vertex, which plays an important role in the chiral symmetry breaking, we consider not only its classical tensor structure but also the two most important non-classical tensor structures, namely
\begin{align}
    \left[\Gamma_{A\bar l l}^{(3)}\right]^{a}_{\nu}(p_1,p_2,p_3) &= (2\pi)^4 \delta^{(4)}(p_1+p_2+p_3)\nonumber   \\[2ex]  
	&\quad\times\left[Z_{A}^{\frac{1}{2}}(p_1) Z_{l}^{\frac{1}{2}}(p_2)Z_{l}^{\frac{1}{2}}(p_3)\right]\sum_{i=1,4,7}\lambda_{A\bar l l}^{(i)}(\bar{p})\Pi_{\mu\nu}^{\perp}(p_1) \left[ {\cal T}_{A\bar l l}^{(i)}(p_1,p_2,p_3)\right]^{a}_{\mu} \,,\label{eq:quark-gluon-Gamma}
\end{align}
with
\begin{align}
    &\left[ {\cal T}_{A\bar l l}^{(1)}(p_1,p_2,p_3)\right]^{a}_{\mu} = T_{c}^{a} \left(\mathrm{i}\gamma_\mu\right) \,,\quad \left[ {\cal T}_{A\bar l l}^{(4)}(p_1,p_2,p_3)\right]^{a}_{\mu} = T_{c}^{a} \left(\mathrm{i}\sigma_{\mu\alpha}p_{1,\alpha}\right) \,,\nonumber   \\[2ex]
    &\left[ {\cal T}_{A\bar l l}^{(7)}(p_1,p_2,p_3)\right]^{a}_{\mu} = T_{c}^{a} \left[\frac{1}{3}\Big(\sigma_{\alpha\beta}\gamma_{\mu}+\sigma_{\beta\mu}\gamma_{\alpha}+\sigma_{\mu\alpha}\gamma_{\beta}\Big) (p_1+ p_2)_{\alpha}(p_1- p_2)_{\beta}\right] \,,\label{eq:quark-gluon-channel}
\end{align}
where $T_{c}^{a}$ are the generators of the gauge group in the fundamental representation, and the Dirac tensor $\sigma_{\mu\nu}=\frac{1}{2}[\gamma_\mu,\gamma_\nu]$. The non-classical tensor structures ${\cal T}_{A\bar l l}^{(4,7)}$ make sizable contributions to the self-consistent chiral symmetry breaking in QCD, while the contributions from other non-classical tensor structures are negligible, see \cite{Fu:2025hcm, Mitter:2014wpa, Williams:2014iea, Williams:2015cvx,  Aguilar:2016lbe, Cyrol:2017ewj, Gao:2021wun, Ihssen:2024miv}. In particular, ${\cal T}_{A\bar l l}^{(4)}$ breaks the chiral symmetry, whereas ${\cal T}_{A\bar l l}^{(7)}$ is chirally symmetric. Furthermore, we also consider the non-classical two-light-quark-two-gluon vertex, namely
\begin{align}
    \left[\Gamma_{AA\bar l l}^{(4)}\right]^{a_1 a_2}_{\mu_3\mu_4}(p_1,p_2,p_3,p_4) = &(2\pi)^4 \delta^{(4)}(p_1+p_2+p_3+p_4)\times\left[Z_{A}^{\frac{1}{2}}(p_1) Z_{A}^{\frac{1}{2}}(p_2)Z_{l}^{\frac{1}{2}}(p_3)Z_{l}^{\frac{1}{2}}(p_4)\right]  \nonumber   \\[2ex]  
    &\times\Pi_{\mu_1\mu_3}^{\perp}(p_1)\Pi_{\mu_2\mu_4}^{\perp}(p_2)\lambda_{AA\bar l l}(\bar{p})\left[ {\cal T}_{AA\bar l l}(p_1,p_2,p_3,p_4)\right]^{a_1 a_2}_{\mu_1\mu_2} \,,
\end{align}
Here we choose the tensor structure that makes the largest contribution to the chiral symmetry breaking, see \cite{Fu:2025hcm,Mitter:2014wpa,Cyrol:2017ewj,Ihssen:2024miv},
\begin{align}
    \left[ {\cal T}_{AA\bar l l}(p_1,p_2,p_3,p_4)\right]^{a_1 a_2}_{\mu_1\mu_2} = \delta_{\mu_1\mu_2}\big\{T_{c}^{a_1},T_{c}^{a_2}\big \}+2\sigma_{\mu_1\mu_2}f^{a_1 a_2 b}T_{c}^{b} \,,
\end{align}
According to the gauge-consistent approximation, the dressing of this vertex can be approximated as
\begin{align}
    \lambda_{AA\bar l l}(\bar{p})\approx \frac{1}{\sqrt{\bar p^2}} \lambda_{A\bar l l}^{(4)}(\bar{p})\,.
\end{align}
For the strange quark-gluon vertex, the effective action is truncated to only the classical tensor structure ${\cal T}_{A\bar s s}^{(1)}$.

For the three-gluon, four-gluon, ghost-gluon, and quark-gluon vertices appearing in the classical action, we define the QCD strong couplings as
\begin{align}
    \alpha_{A^{3}}(p)&=\frac{\lambda^2_{A^{3}}(p)}{4\pi}\,,\quad \alpha_{A^{4}}(p)=\frac{\lambda_{A^{4}}(p)}{4\pi}\,,\quad \alpha_{A\bar c c}(p)=\frac{\lambda^2_{A\bar c c}(p)}{4\pi}\,,\\[2ex]
    \alpha_{A\bar l l}(p)&=\frac{\lambda^2_{A\bar l l}(p)}{4\pi}\,,\quad
    \alpha_{A\bar s s}(p)=\frac{\lambda^2_{A\bar s s}(p)}{4\pi}\,.\label{eq:strong-coupling}
\end{align}
In the ultraviolet perturbative momentum region, all the strong couplings should agree with each other.

The four-quark vertex in the pure matter part dominates the breaking of chiral symmetry in the low-energy region and the emergence of meson resonance states. The vertex dressing contains information about the meson pole mass and the Bethe-Salpeter amplitude. The four-quark vertex can be explicitly expressed as
\begin{align}
    \left[\Gamma_{4q}^{(4)}\right]_{ijkm}^{(\alpha)}&=-(2\pi)^4 \delta^{(4)}(\sum_{i=1}^{4}p_i)\times\left[\prod_{i=1}^4 Z_{q}^{\frac{1}{2}}(p_i)\right]  \nonumber\\[2ex]&\quad\times \lambda_{\alpha}(p_1,p_2,p_3,p_4)\left[ {\cal T}_{\alpha}(p_1,p_2,p_3,p_4)\right]_{ijkm} \,.
    \label{eq:4q-lambda}
    \end{align}
Here, we consider the four most important four-quark interaction channels in vacuum $\{\sigma,\pi,\kappa,K\}$, which are discussed in \cite{Fu:2022uow,Fu:2024ysj,Fu:2025hcm},
\begin{align}
    &{\cal T}_{4q,ijkm}^{\sigma} \bar q_{i} q_{j} \bar q_{k} q_{m} =\left(\bar{q}\,T^{0}q\right)^2\,,\quad {\cal T}_{4q,ijkm}^{\pi} \bar q_{i} q_{j} \bar q_{k} q_{m} =-\left(\bar{q}\,\gamma_5 T^{(1-3)}q\right)^2\,,\nonumber   \\[2ex]
    &{\cal T}_{4q,ijkm}^{\kappa} \bar q_{i} q_{j} \bar q_{k} q_{m} =\left(\bar{q}\,T^{(4-7)}q\right)^2\,,\quad {\cal T}_{4q,ijkm}^{K} \bar q_{i} q_{j} \bar q_{k} q_{m} =-\left(\bar{q}\,\gamma_5 T^{(4-7)}q\right)^2\,,
\end{align}
where $T$ are the Gell-Mann matrices in the flavor space, and $q=(l,l,s)$ denotes the 2+1 flavour quark field.

For the dressing $\lambda_{\alpha}$, we first introduce $s,t,u$-channel momentum approximation, and then present a concrete procedure to define and extract the meson pole masses and Bethe–Salpeter amplitudes from the dressing. With this truncation of momenta, the dressing can be approximated as
\begin{align}
    \lambda_{\alpha}(p_1,p_2,p_3,p_4)= \lambda_{\alpha}(s,t,u)+\Delta\lambda_{\alpha}(p_1,p_2,p_3,p_4)\approx \lambda_{\alpha}(s,t,u) \,.
\end{align}
It is convenient to adopt the momenta, as follows
\begin{align}
    P=-(p_1+p_2)\,,\quad p=\frac{p_2-p_1}{2}\,,\quad p'=\frac{p_4-p_3}{2}\,,
\end{align}
such that the Mandelstam variables $s$, $t$, and $u$ read
\begin{align}
    s=(p+p')^2\,,\quad t=P^2\,,\quad u=(p-p')^2\,,
\end{align}
In order to extract the meson pole mass and Bethe–Salpeter amplitude from the dressing, we take specifically the momentum configuration as follows
\begin{align}
P_{\mu}=\sqrt{P^2}(1,0,0,0)\,,\quad p_{\mu}=|p|(\cos\theta,\sin\theta,0,0)\,,\quad p'_{\mu}=-|p|(\cos\theta,\sin\theta,0,0)\,.\label{eq:momentum-configuration}
\end{align}
Consequently, the dressing can be equivalently expressed as 
\begin{align}
    \lambda_{\alpha}(s,t,u)= \lambda_{\alpha}(P,p,p')= \lambda_{\alpha}(P^2,|p|,\cos\theta) \,.
\end{align}

Mesons, as the lowest-lying resonances in the respective four-quark channels, have pole masses that correspond to the first singularity of the four-quark vertex dressing in the $t$-channel in Minkowski space. Taking the pion studied in this work as an example, its pole mass $m_{\pi}$ satisfies
\begin{align}
    \frac{1}{\lambda_{\pi}(t=-m_{\pi}^2)}=0\,,\label{eq:pole-pion}
\end{align}
In the numerical calculations, we use the Pad\'e approximation to analytically continue the dressing data of $t>0$. For the extraction of the first pole at $t<0$, this method is very stable and reliable, see \cite{Fu:2022uow,Fu:2024ysj,Fu:2025hcm}. Furthermore, with the momentum configuration in \labelcref{eq:momentum-configuration}, the pion BS amplitude can be extracted from the residue of the four-quark vertex dressing at the pole of the pion bound state,
\begin{align}
    h_{\pi}(p,\cos\theta)= \lim_{P^2\to -m_{\pi}^2} \sqrt{\lambda_{\pi}(P,p,\cos\theta)\cdot(P^2+m_{\pi}^2)} \,,
\end{align}
Employing the BS amplitude and the quark two-point function, one is able to compute meson-related observables such as the decay constants and PDAs.

Within this framework, all correlation functions are computed via the fRG method and fed back into the flow equations, yielding a self-contained and self-consistent first-principles QCD calculation. With a small strong coupling $\alpha_{s,\Lambda}$ at the chosen UV cutoff scale $\Lambda$, the only input parameters are the light and strange running quark masses $m_l$ and $m_s$. Under the combined effects of quark-gluon interactions and four-quark interactions, the system realises quantitative chiral symmetry breaking. In the calculation, the parameters are chosen as
\begin{align}
    \alpha_{s,\Lambda}=0.179\,\,\,\mathrm{with}\,\,\,\Lambda=35.7\text{ GeV}\,,\quad m_l=2.1\text{ MeV}\,,\
    \quad m_s=55.9\text{ MeV}\,,
\end{align}
where the running quark masses are fixed by the ratios of physical observables,
\begin{align}
    \frac{m_{\pi}}{f_{\pi}}=\frac{137\text{ MeV}}{93\text{ MeV}}\,,\quad \frac{m_{K}}{f_{\pi}}=\frac{494\text{ MeV}}{93\text{ MeV}}\,,
\end{align}
where the pion decay constant is computed from
\begin{align}
    \mathrm{i}P_{\mu}f_{\pi}\delta^{ab}=\bra{0}J_{5\mu}^{a}\ket{\pi^{b}}=\frac{\delta^{ab}}{2}\int\frac{d^4 q}{(2\pi)^4}\mathrm{Tr}\Big[\gamma_\mu \gamma_5 G_{l}(q+P)h_{\pi}(q)\gamma_5 G_{l}(q)\Big]\,,
\end{align}
with the light quark propagator $G_{l}$. In addition, all other correlation functions and observables are theoretical predictions, such as
\begin{align}
    m_\sigma = 515.2 \,\mathrm{MeV}\,,\quad
    f_K = 114.1 \, \mathrm{MeV}\,,\quad
    M_l = 344.5 \, \mathrm{MeV}\,,\quad
    M_s = 487.3 \, \mathrm{MeV}\,,
\end{align}
where $M_l=M_l(p=0)$ and $M_s=M_s(p=0)$ are the constituent quark masses for the light and strange quarks, respectively, $f_K$ is the kaon decay constant, and $m_\sigma$ is the pole mass of the $\sigma$-mode.

%
\begin{figure}[t]
    \includegraphics[width=0.45\textwidth]{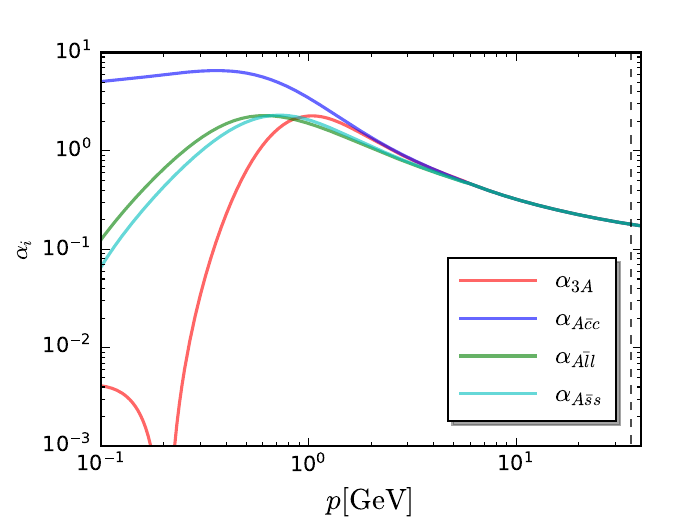}\hspace{0.5cm}
    \includegraphics[width=0.45\textwidth]{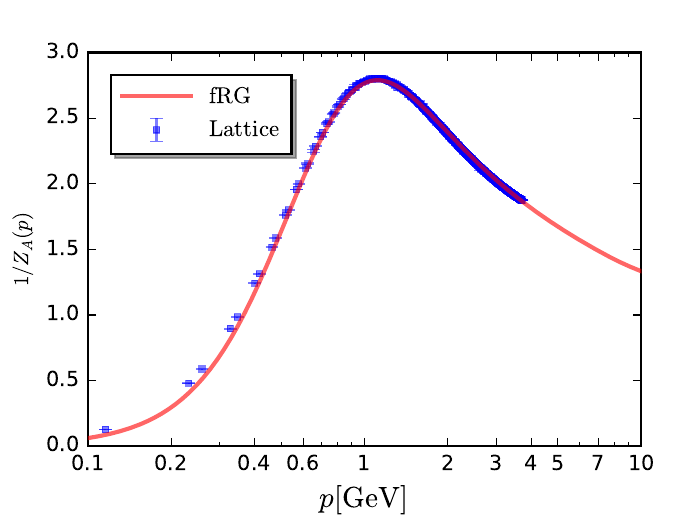}
    \caption{Left panel: QCD strong couplings $\alpha_{i}$ for $i=A^{3},A\bar c c,A\bar l l,A\bar s s$ as a function of the symmetric point momentum $p$, which are defined from the vertex dressings in \Cref{eq:strong-coupling}. Right panel: Gluon propagator dressing $1/Z_{A}(p)$ as a function of the momentum $p$ in functional QCD in comparison to the unquenched lattice QCD results for $N_{f}=2+1$ flavours \cite{Boucaud:2018xup}. }\label{fig:glue}
\end{figure}
%

In the main text, we have presented results of the quark two-point function and the pion BS amplitude, directly used to compute the pion PDA. Here we close this appendix with results for other important QCD correlation functions; more results can be found in \cite{Fu:2025hcm}. In the left panel of \Cref{fig:glue}, the strong couplings defined from different vertices show good agreement in the perturbative and semi-perturbative momentum regime, $p\gtrsim 3$\,GeV. In the low-momentum non-perturbative regime, $p\lesssim 3$\,GeV, however, they gradually deviate significantly from each other due to the enhancement of the strong couplings and the emergence of the gluon confinement mass gap; see also \cite{Aguilar:2011xe, Cyrol:2016tym, Ferreira:2025anh}. In this computation, we take the approximation $\alpha_{A^{3}}(p)=\alpha_{A^{4}}(p)$.

In the right panel of \Cref{fig:glue}, we show the gluon propagator dressing featuring a confinement mass gap. The functional results are in good agreement with the lattice QCD results of \cite{Boucaud:2018xup}. Due to the mass gap, the gluon decouples from the system at scales below about 300 MeV, see also \cite{Ihssen:2024miv}.

%
\begin{figure}[t]
    \includegraphics[width=0.45\textwidth]{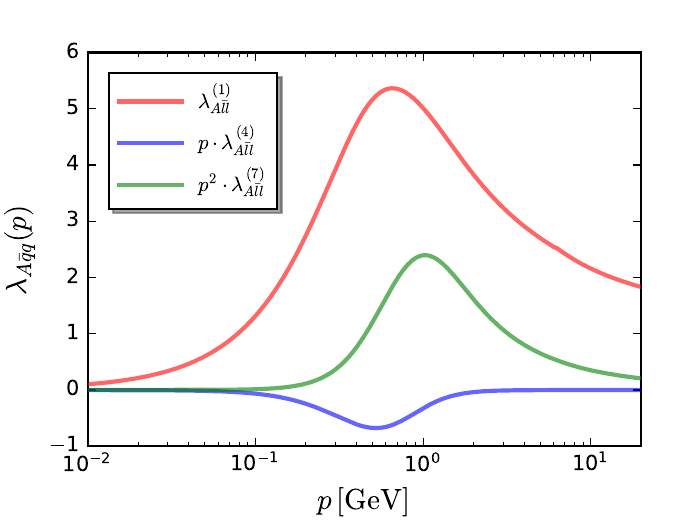}\hspace{0.5cm}
    \includegraphics[width=0.45\textwidth]{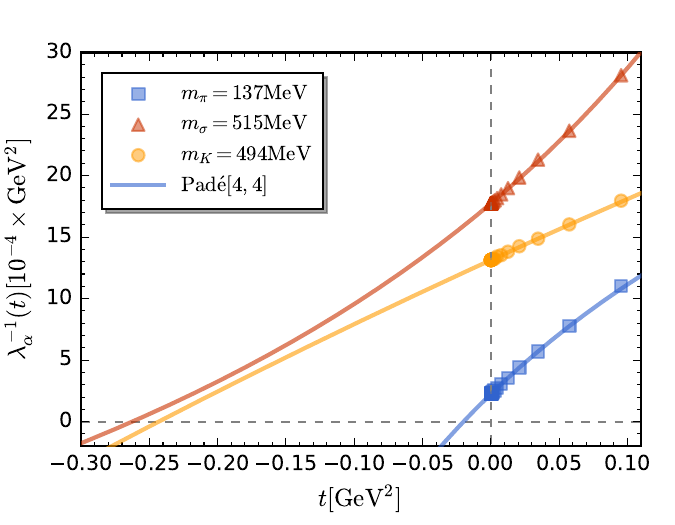}
\caption{Left panel: Dimensionless light quark-gluon vertex dressings $\lambda^{(1,4,7)}_{A\bar l l}(p)$ as a function of the symmetric point momentum $p$ for the tensor structures ${\cal T}^{(1,4,7)}_{A\bar l l}$ in \Cref{eq:quark-gluon-channel}. Right panel: Inverse four-quark coupling $1/\lambda_{\alpha}(t)$ for $\alpha = \pi, \sigma, K$ as a function of the $t$-channel momentum. The data points denote Euclidean results, and the solid lines are Pad\'e[4,4] fits. The pole masses are determined by the $t$-channel value where $1/\lambda_{\alpha}(t)=0$, see \Cref{eq:pole-pion}.}\label{fig:quark}
\end{figure}
%

For the glue-matter interface term, the left panel of \Cref{fig:quark} shows the dressing results for the classical and non-classical quark-gluon vertex channels in \Cref{eq:quark-gluon-Gamma,eq:quark-gluon-channel}. Finally, in the right panel of \Cref{fig:quark}, we show the inverse four-quark vertex dressing as a function of the $t$-channel momentum. Moreover, in this plot we also show the result of analytic continuation for $\lambda_{\alpha}^{-1}(t)$ using the fourth-order Pad\'e approximation, denoted by $\text{pad\'e}[4,4]$. One can extract the pole mass and Bethe-Salpeter amplitude of the corresponding bound state at  $\lambda_{\alpha}^{-1}(t)=0$. This provides the key foundation for computing the PDA in this work.

%
\begin{figure}[t]
\includegraphics[width=0.45\textwidth]{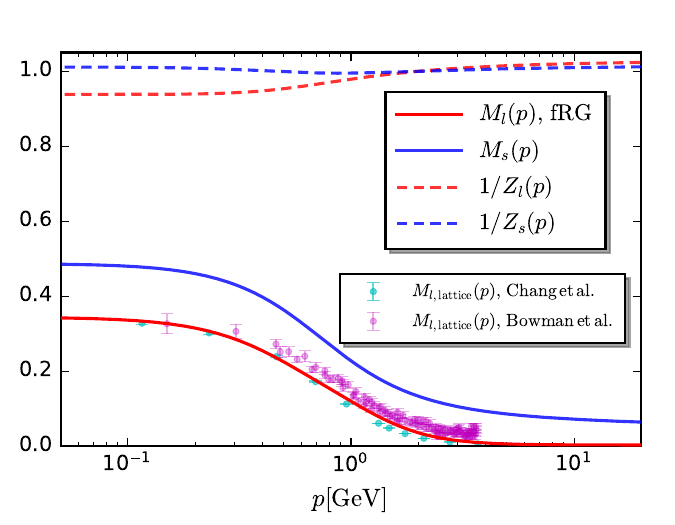}
\caption{Quark mass function $M_{q}(p)$ (solid line, GeV) and the quark wave function $Z_{q}(p)$ (dashed line) for the light (red line) and strange quark (blue line). The lattice data are taken from \cite{Chang:2021vvx}  (cyan points) and \cite{Bowman:2005vx}  (purple points).}\label{fig:Quark-Nf2+1}
\end{figure}
%

In \Cref{fig:Quark-Nf2+1}, we show the quark mass function and wave function. In this 2+1 flavour QCD computation, the two-point correlation function of the strange quark is also solved self-consistently.

\subsection{Deformed integration contour in quasi-PDA calculations}
\label{sec:integrate-contour}

In \Cref{eq:qPDA}, we have presented the definition of the quasi-PDA. This integral can be expanded more explicitly as
\begin{align}
    \phi_{\pi}(x,P_z)&=\frac{1}{f_{\pi}}\frac{4N_c}{(2\pi)^4}\int dp_{\perp}^{2}dp_{0}\,h_{\pi}(p,\cos\theta)P_{z}\frac{1}{Z_{l}^{1/2}(p_{+})Z_{l}^{1/2}(p_{-})}\frac{1}{p_{+}^{2}+M_{l}^{2}(p_{+})}\frac{1}{p_{-}^{2}+M_{l}^{2}(p_{-})}
    \nonumber\\[2ex]
    &\quad\times\left[xM_{l}(p_{-})+(1-x)M_{l}(p_{+})\right]\,,\label{eq:qPDA-expand}
\end{align}
where one has
\begin{align}
    p_{\pm}=p\pm\frac{P}{2}\,,\quad \text{with}\,\,\,p_{\mu}=(p_{0},p_{3},p_{\perp})\,\,\,\text{and}\,\,\,P_{\mu}=(\mathrm{i}E_{\pi},P_{z},0,0)\,.\label{eq:p-P}
\end{align}
and
\begin{align}
p_{3}=\left(x-\frac{1}{2}\right)P_{z}\,,\quad \cos\theta=\frac{p_{0}}{p}\,,\quad p=\sqrt{p_{0}^{2}+p_{3}^{2}+p_{\perp}^{2}}\,.
\end{align}

As shown in \Cref{eq:qPDA-expand}, in the functional quasi-PDA framework used in this work, although the integration over the internal momentum $p$ is performed in Euclidean space, the choice of $P$ in \Cref{eq:p-P} requires analytic continuation of the integral into the complex plane, and the region that must be continued grows as $P_z$ increases. Moreover, as $P_z$ increases, the pole structure in \Cref{eq:qPDA-expand} is also distorted. The two poles in \Cref{eq:qPDA-expand} for the quark read
\begin{align}
    &p_{0,\mathrm{pole}\,1}=\mathrm{i}\left[-\frac{1}{2}\sqrt{P_{z}^{2}+m_{\pi}^{2}}-\sqrt{p_{\perp}^{2}+(xP_{z})^2+M_{l}^2}\right]\,,\nonumber\\[2ex] &p_{0,\mathrm{pole}\,2}=\mathrm{i}\left[-\frac{1}{2}\sqrt{P_{z}^{2}+m_{\pi}^{2}}+\sqrt{p_{\perp}^{2}+(xP_{z})^2+M_{l}^2}\right]\,,\label{eq:pole-quark}
\end{align}
For the antiquark, its two poles are given by
\begin{align}
    &p_{0,\mathrm{pole}\,3}=\mathrm{i}\left[\frac{1}{2}\sqrt{P_{z}^{2}+m_{\pi}^{2}}-\sqrt{p_{\perp}^{2}+\left((1-x)P_{z}\right)^2+M_{l}^2}\right]\,,\nonumber\\[2ex] & p_{0,\mathrm{pole}\,4}=\mathrm{i}\left[\frac{1}{2}\sqrt{P_{z}^{2}+m_{\pi}^{2}}+\sqrt{p_{\perp}^{2}+\left((1-x)P_{z}\right)^2+M_{l}^2}\right]\,,\label{eq:pole-antiquark}
\end{align}
For vanishing and small $P_z$, $p_{0,\mathrm{pole}\,1}$ and $p_{0,\mathrm{pole}\,3}$ always lie in the lower half of the complex plane, while $p_{0,\mathrm{pole}\,2}$ and $p_{0,\mathrm{pole}\,4}$ always lie in the upper half. This pole distribution is the same as that on the light cone. As $P_z$ increases, $p_{0,\mathrm{pole}\,2}$ crosses the real axis into the lower half-plane in some $x$ intervals, while $p_{0,\mathrm{pole}\,3}$ crosses the real axis into the upper half-plane in the corresponding $(1-x)$ intervals. This unphysical artifact introduced by finite $P_z$ renders the quasi-PDA calculation invalid.

To resolve this issue, we develop a deformed integration contour method within this framework \cite{Zhang:2025ofc}, namely shifting the integration by a finite imaginary part, i.e.,
\begin{align}
    \int_{-\infty}^{\infty} dp_{0} \rightarrow \int_{-\infty+\mathrm{i}C}^{\infty+\mathrm{i}C} dp_{0} \,,
\end{align}
To avoid altering the distribution of $p_{0,\mathrm{pole}\,1}$ and $p_{0,\mathrm{pole}\,4}$, we choose $C$ as
\begin{align}
    C=\frac{\mathrm{Im}(p_{0,\mathrm{pole}\,2})+\mathrm{Im}(p_{0,\mathrm{pole}\,3})}{2}\,.
\end{align}
In this way, even in numerical integration, we can ensure that the quark and antiquark pairs $(p_{0,\mathrm{pole}\,1},p_{0,\mathrm{pole}\,3})$ or $(p_{0,\mathrm{pole}\,2},p_{0,\mathrm{pole}\,4})$ are taken into account simultaneously.

Finally, the imaginary shift in the $p_{0}$ direction is only introduced to adjust the pole distribution during integration. Therefore, as long as the analytic structure of the quark--antiquark pair is unchanged, the choice of $C$ does not introduce numerical arbitrariness even when the BS amplitude and quark masses are momentum dependent. For more details, see \cite{Zhang:2025ofc}, where detailed analytic and numerical discussions and examples are provided.

\subsection{Taylor expansion of the Bethe-Salpeter amplitude and quark mass in the complex plane}
\label{sec:expansion}

With the contour-shift method, the present calculation still has an upper limit on $P_z$. This limitation arises from the relative positions of $p_{0,\mathrm{pole}\,2}$ and $p_{0,\mathrm{pole}\,3}$ in the complex plane in \Cref{eq:pole-quark,eq:pole-antiquark}, namely their imaginary parts must satisfy
\begin{align}
    \mathrm{Im}(p_{0,\mathrm{pole}\,2})-\mathrm{Im}(p_{0,\mathrm{pole}\,3})>0\,.
\end{align}
For the pion, since both the quark and antiquark are light quarks, the above difference is minimal at $x=1/2$. Since this constraint must hold over the entire integration region, it can be finally reduced to
\begin{align}
    M_{l}(p)>m_{\pi}/2\,.
\end{align}
Physically, this limitation follows from the stability condition for the formation of meson; otherwise the bound state would dissolve into a quark and antiquark pair. For fixed physical quark mass and pion mass, this condition is always satisfied, but for momentum-dependent quark mass functions in Euclidean space, it will be violated at large $P_z$.

In order to get access to larger $P_z$ values in the quasi-PDA calculation, we need information on the quark mass function in the complex plane. Moreover, as discussed in \Cref{sec:integrate-contour}, the quasi-PDA calculation requires analytic continuation of the integral into the complex plane. It would improve the accuracy of calculations by including (or partially including) the analytic information of the quark wave function and the pion BS amplitude in the complex plane. However, as shown in \Cref{fig:Quark,fig:BS-amplitude} and discussed in \Cref{sec:2+1QCD}, the direct functional QCD calculation provides the quark two-point function and pion BS amplitude only for Euclidean momentum $p>0$. Here we adopt a Taylor expansion in the complex plane to improve the calculation. After shifting the integration momentum, the momentum of the pion BS amplitude is
\begin{align}
    p_{\mu}=\Big(p_{0}+\mathrm{i}\,C,\,(x-1/2)P_{z},\,p_{\perp}\Big)\,,
\end{align}
The quark and antiquark momenta read
\begin{align}
     p_{+\mu}&=\Big(p_{0}+\mathrm{i}\,\big(C+\frac{\sqrt{P_{z}^{2}+m_{\pi}^{2}}}{2}\big),\,x P_{z},\,p_{\perp}\Big)\,,\\[2ex]
     p_{-\mu}&=\Big(p_{0}+\mathrm{i}\,\big(C-\frac{\sqrt{P_{z}^{2}+m_{\pi}^{2}}}{2}\big),\,(x-1)P_{z},\,p_{\perp}\Big)\,.
\end{align}
Consequently, the Taylor expansion in the imaginary part can be written as
\begin{align}
    h_{\pi}(p,\cos\theta)= h_{\pi}(\bar p,\cos\theta)+\sum_{k=1}^{n}\frac{1}{k!}\left(\frac{\partial}{\partial p_{0}}\right)^k h_{\pi}(\bar p,\cos\theta)\left(\mathrm{i}\,C\right)^k+\mathcal{O}(\left(\mathrm{i}\,C\right)^{n+1})\,,\label{eq:Taylor-expansion-h}
\end{align}
with $\bar p_{\mu}=(p_{0},\,(x-1/2)P_{z},\,p_{\perp})$.

Similarly, the quark mass functions can be expanded as
\begin{align}
    M_{l}(p_{+}) &= M_{l}(\bar{p}_{+})+\sum_{k=1}^{n}\frac{1}{k!}\left(\frac{\partial}{\partial p_{0}}\right)^k M_{l}(\bar{p}_{+})\left(\Delta p_{0}^{+}\right)^k+\mathcal{O}(\left(\Delta p_{0}^{+}\right)^{n+1})\,,\nonumber\\[2ex]
    M_{l}(p_{-}) &= M_{l}(\bar{p}_{-})+\sum_{k=1}^{n}\frac{1}{k!}\left(\frac{\partial}{\partial p_{0}}\right)^k M_{l}(\bar{p}_{-})\left(\Delta p_{0}^{-}\right)^k+\mathcal{O}(\left(\Delta p_{0}^{-}\right)^{n+1})\,,\label{eq:Taylor-expansion-M}
\end{align}
with 
\begin{align}
    \bar{p}_{+\mu}=(p_{0},\,xP_{z},\,p_{\perp})\,,\quad \bar{p}_{-\mu}=(p_{0},\,(x-1)P_{z},\,p_{\perp})\,.
\end{align}
and
\begin{align}
    \Delta p_{0}^{+}=\mathrm{i}\,\left(C+\frac{\sqrt{P_{z}^{2}+m_{\pi}^{2}}}{2}\right)\,,\quad \Delta p_{0}^{-}=\mathrm{i}\,\left(C-\frac{\sqrt{P_{z}^{2}+m_{\pi}^{2}}}{2}\right)\,.
\end{align}
Following the same procedure, the quark wave functions $Z_{l}(p_{+})$ and $Z_{l}(p_{-})$ are also expanded. 

%
\begin{figure}[t]
    \includegraphics[width=0.45\textwidth]{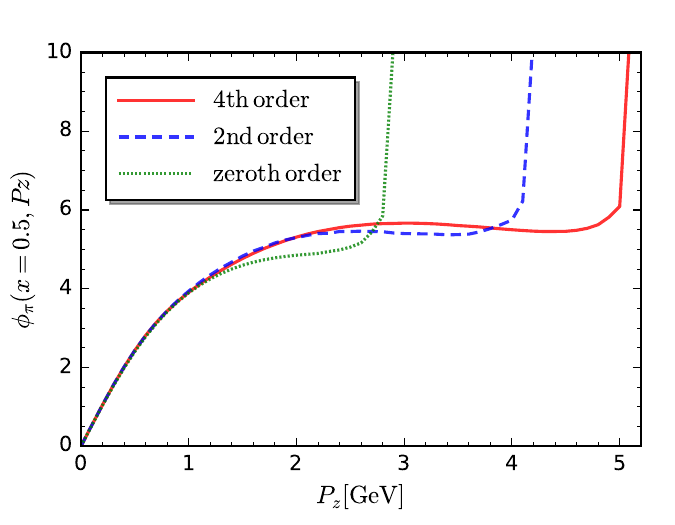}\hspace{0.5cm}
    \includegraphics[width=0.45\textwidth]{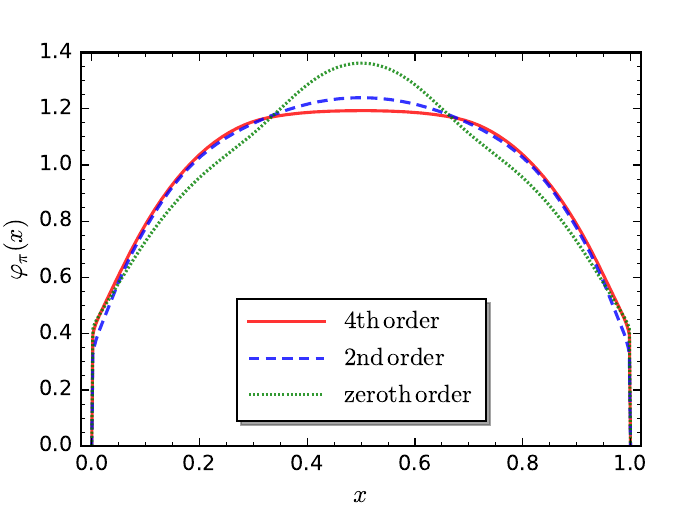}
    \caption{Left panel: Non-normalised quasi-PDA $\phi_{\pi}(x,P_{z})$ at $x=0.5$ as a function of the longitudinal momentum $P_{z}$. The lines of different colors show the quasi-PDA obtained with the zeroth-, second-, and fourth-order Taylor expansions of the light-quark mass function and BS amplitude, respectively, see \Cref{eq:Taylor-expansion-h,eq:Taylor-expansion-M}. Right panel: Normalised pion PDA $\varphi_{\pi}(x)$ as a function of the momentum fraction $x$. The lines of different colors show the PDA obtained with zeroth-, second-, and fourth-order Taylor expansions of the light-quark mass function and BS amplitude, see \Cref{eq:Taylor-expansion-h,eq:Taylor-expansion-M}.}
    \label{fig:Taylor-expansion}
\end{figure}
%

Due to the numerical precision and the convergence of the Taylor expansion in the complex plane, we find that expanding to the fourth order is sufficient to yield results with a larger $P_{z}$ range, better convergence, and numerical stability. The left panel of \Cref{fig:Taylor-expansion} shows the quasi-PDA at $x=0.5$ as a function of $P_{z}$ for different expansion orders. Compared with the result without expansion, the fourth-order expansion extends the accessible $P_{z}$ range to more than twice that of the zeroth-order result, up to $P_{z}\simeq 4.5$ GeV. Moreover, in the fourth-order results we observe a clear plateau of the quasi-PDA in the range $2.5 \,\mathrm{GeV} \lesssim P_{z} \lesssim 4.5\,\mathrm{GeV} $, indicating that the saturation of quasi-PDA with the increase of $P_{z}$ and the convergence are observed, which is very suited for the use of LaMET extrapolation. This conclusion is also verified in the results in \Cref{fig:quasi-PDA} and \Cref{fig:Pz-x-extrap} in \Cref{sec:extrapolation}.

In the right panel of \Cref{fig:Taylor-expansion}, we present the final PDA results after extrapolation for different orders of the Taylor expansion. Compared with the fourth-order result, the zeroth-order result shows a sizable deviation; the second-order result agrees with the fourth-order one except for a small deviation in the central region $0.35<x<0.65$. This indicates that as the expansion order increases, more information of the correlation functions in the complex plane is incorporated, and the PDA converges to a stable result. 


\subsection{large $P_{z}$ extrapolation and boundary $x$ extrapolation}
\label{sec:extrapolation}

%
\begin{figure}[t]
    \includegraphics[width=0.45\textwidth]{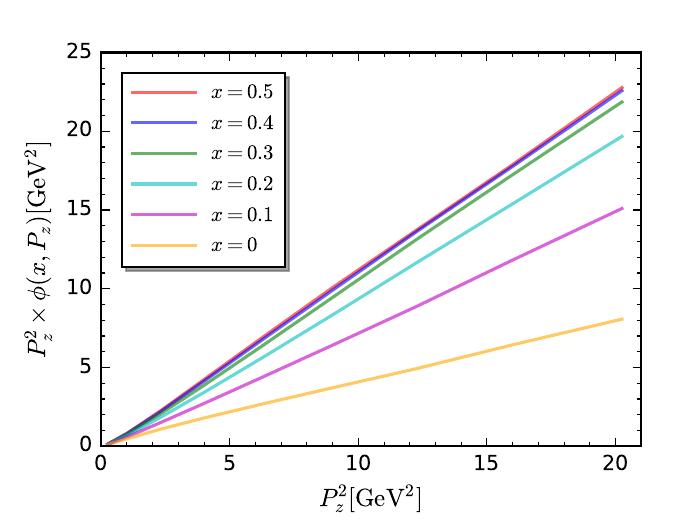}
    \hspace{0.5cm}
    \includegraphics[width=0.45\textwidth]{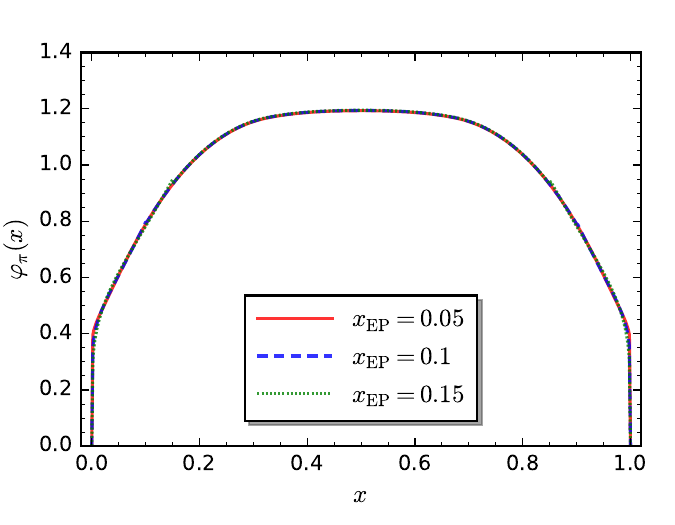}\caption{Left panel: $P_{z}^{2}\phi_{\pi}(x,P_{z})$ as a function of $P_{z}^{2}$. Lines of different colors denote the results for different $x$ values. Right panel: Normalised pion PDA $\varphi_{\pi}(x)$ as a function of the momentum fraction $x$. The different lines show the PDA obtained with different choices of the endpoint fitting interval in momentum fraction, $x_{\mathrm{EP}}=0.05, 0.1, 0.15$, see \Cref{eq:x-extrap}.}
    \label{fig:Pz-x-extrap}
\end{figure}
%

\begin{table}[t]
  \centering
  \begin{tabular}{c|ccc}
    \hline\hline & &    \\[-2ex]   
       & a & b & c \\[1ex]
      \hline & & &  \\[-2ex]
      \,\,$x_{\mathrm{EP}}=0.05$\,\, & \,\,0.047\,\, &  \,\,-5.694\,\,  &  \,\,0.500\,\, \\[1ex]
      \hline & &   \\[-2ex]
      \,\,$x_{\mathrm{EP}}=0.1$\,\, & \,\,0.113\,\, &  \,\,-3.547\,\,  &  \,\,0.680\,\,   \\[1ex]
      \hline & &   \\[-2ex]
      \,\,$x_{\mathrm{EP}}=0.15$\,\, & \,\,0.184\,\, &  \,\,-2.040\,\,  &  \,\,0.921\,\,    \\[1ex]
      \hline\hline
  \end{tabular}
  \caption{Fitting parameters for different $x$ fitting ranges $0<x<x_{\mathrm{EP}}$ shown in \Cref{eq:x-extrap}.}
  \label{tab:x-fit}
\end{table}

According to the LaMET \cite{Ji:2013dva, Ji:2020ect}, the light-cone PDA can be obtained by extrapolating the quasi-PDA at large $P_{z}$. In contrast to the matching relation in lattice QCD approach \cite{Ji:2015qla, Zhang:2017bzy, LatticeParton:2022zqc}, the matching relation between the quasi-DA and the light-cone DA in the functional framework is affected only by higher-twist corrections without the perturbative terms. This follows from two observations. First, with the nonperturbative quark propagator and BS amplitude shown in \Cref{fig:Quark,fig:BS-amplitude} as inputs, the quasi-DA integral is ultraviolet finite and does not require the introduction of an additional ultraviolet cutoff. Second, as discussed in \Cref{sec:2+1QCD}, the conversion between different renormalisation schemes amounts only to an overall multiplicative factor. Since both the LFWF and the DA are normalized, a constant factor does not affect the final results. Therefore, this extrapolation can be written as
\begin{align}
    \phi_{\pi}(x,P_z)= \phi_{\pi}(x,P_z \to \infty)+ \frac{c_{2}(x)}{P_{z}^{2}} +\mathcal{O}(\frac{1}{P_{z}^{4}}) \,,\label{eq:LaMET-extrap}
\end{align}
Here $c_{2}(x)$ is the extrapolation coefficient. In this work, we find that expanding to order $1/P_{z}^{2}$ is sufficient to obtain stable extrapolation results. In this form, \Cref{eq:LaMET-extrap} can be rewritten as
\begin{align}
    P_{z}^{2}\phi_{\pi}(x,P_z)\simeq P_{z}^{2}\phi_{\pi}(x,P_z \to \infty)+ c_{2}(x) \,.
\end{align}
Therefore, a successful extrapolation implies that $P_{z}^{2}\phi_{\pi}(x,P_z)$ depends linearly on $P_{z}^{2}$. The left panel of \Cref{fig:Pz-x-extrap} shows $P_{z}^{2}\phi_{\pi}(x,P_z)$ as a function of $P_{z}^{2}$ at different $x$ values. One sees a clear linear trend consistent with this expectation. The coefficient of $1/P_{z}^{2}$ term $c_{2}(x)$ is around $0.2-0.3\, \text{GeV}^2$  in the whole $x$ regime, which is small and also supported in the left panel of \Cref{fig:Pz-x-extrap}. 

On the other hand, the LaMET is not applicable near the endpoints $x=0$ and $x=1$, see \cite{Ji:2020ect}. In \Cref{fig:quasi-PDA}, the quasi-PDA is not strictly zero at $x=0$ and $x=1$, but it rapidly decreases to zero outside the physical region. Specifically, to obtain the final PDA, we fit the endpoint regions in $x$ based on $\phi_{\pi}(x,P_z \to \infty)$ with a phenomenological function, see \cite{LatticeParton:2022zqc},
\begin{align}
    \varphi_{\pi}(x)= c\,x^{a}(1-x)^{b}\,,\,\,\,\text{for}\,\,\,0<x<x_{\mathrm{EP}}\,\,\,\,\text{and}\,\,\,1-x_{\mathrm{EP}}<x<1\,,\label{eq:x-extrap}
\end{align}
where $x_{\mathrm{EP}}$ sets the size of the endpoint region, $a$, $b$ and  $c$ are the shape parameters of the fit function, and $\varphi_{\pi}(x)$ is the normalised pion PDA.

As shown in \Cref{tab:x-fit}, the fitting parameters in the $x$-endpoint region are not stable due to the limitations of the Euclidean correlation functions inputs and the LaMET approach near the endpoints $x=0$ and $x=1$. However, the behavior in the small-$x$ region has only a minor effect on the PDA at finite $x$. In the right panel of \Cref{fig:Pz-x-extrap}, we show the final PDA results with different choices of the endpoint interval. They coincide with each other with the same moments $\langle\xi^2\rangle_\pi=0.267$, indicating that the extrapolation is insensitive to the choice of $x_{\mathrm{EP}}$.


\end{document}